\documentclass[aps,preprint,prl,superscriptaddress]{revtex4}

\usepackage{graphicx}
\usepackage{amsmath}
\usepackage{amssymb}
\usepackage{units}
\usepackage{color}

\usepackage{setspace}

\begin{document}

\title{Probing carrier dynamics in photo-excited graphene with time-resolved ARPES}

\author{I. Gierz}
\email{Isabella.Gierz@mpsd.mpg.de}
\affiliation{Max Planck Institute for the Structure and Dynamics of Matter, Center for Free Electron Laser Science, Hamburg, Germany}

\date{\today}

\begin{abstract}
The dynamics of photo-generated electron-hole pairs in solids are dictated by many-body interactions such as electron-electron and electron-phonon scattering. Hence, understanding and controlling these scattering channels is crucial for many optoelectronic applications, ranging from light harvesting to optical amplification. Here we measure the formation and relaxation of the photo-generated non-thermal carrier distribution in monolayer graphene with time- and angle-resolved photoemission spectroscopy. Using sub 10fs pulses we identify impact ionization as the primary scattering channel, which dominates the dynamics for the first 25fs after photo-excitation. Auger recombination is found to set in once the carriers have accumulated at the Dirac point with time scales between 100 and 250fs, depending on the number of non-thermal carriers. Our observations help in gauging graphene's potential as a solar cell and TeraHertz lasing material.
\end{abstract}

\maketitle

\section{Keywords}

graphene, ultrafast dynamics, tr-ARPES, Auger scattering

\section{Introduction}

The study of non-equilibrium electronic phenomena impacts many areas of research and technology, from photo-chemical reactions to the efficiency of optoelectronic devices such as lasers \cite{book} and solar cells \cite{Mahan_1979, Palais_2000, Cuevas_2003, Fuyuki_2005}. Furthermore, thermalization and cooling dynamics of non-equilibrium carriers reveal important information about the relevant microscopic many-body interactions \cite{Schmitt_2008, Rohwer_2011, Petersen_2011, Smallwood_2012, Bovensiepen_2012, Hellmann_2012, Sentef_2013}.

In this context, graphene --- a zero-gap semiconductor with a conical band structure \cite{Novoselov_2004, Bostwick_2007} --- is of interest due to its peculiar optical properties. Like in conventional semiconductors, strong optical excitation results in a population-inverted state \cite{Ryzhii_2007, Li_2012, Winzer_2013, Gierz_2013, Gierz_2015_popinv, Jago_2015}.  On the other hand, photo-generated electron-hole pairs in a conventional semiconductor typically relax by Auger recombination \cite{Beattie_1958, Svantesson_1971, Auston_1975, Benz_1976}, a process that is strongly suppressed in photo-excited graphene due to the abence of occupied states at the bottom of the conduction band. Hence, in contrast to conventional semiconductors, primary thermalization events in graphene are dominated by impact ionization \cite{Winzer_2010, Winzer_2012, Plötzing_2014, Gierz_2015_10fs}, where the excess energy of the photo-excited electron is used to generate additional electron-hole pairs.

Here we investigate the buildup and decay of non-thermal carrier distributions in monolayer graphene after photo-excitation. We study these dynamics in different excitation regimes with time- and angle-resolved photoemission spectroscopy (tr-ARPES).
 
\begin{figure}
	\center
  \includegraphics[width = 0.5\columnwidth]{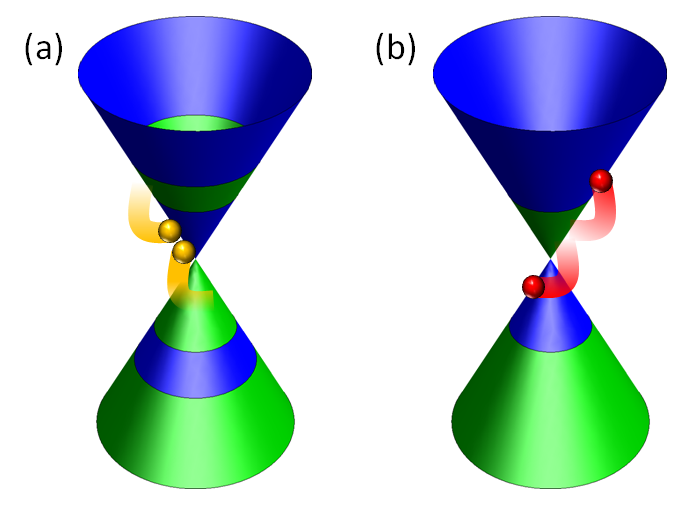}
  \caption{Schematic of the buildup and decay of non-thermal carrier distributions in graphene. (a) Impact ionization accumulates carriers at the bottom of the conduction band. (b) Once the carriers have relaxed to the bottom of the conduction band they decay by Auger recombination.}
  \label{fig1}
\end{figure}

For short pump pulses of durations below 10fs, absorption saturates at relatively low fluences as the density of states around E$_{\text{D}}-\hbar\omega_{\text{pump}}/2$ is depleted and the corresponding states around E$_{\text{D}}+\hbar\omega_{\text{pump}}/2$ are occupied (E$_{\text{D}}$ is the energy of the energy of the Dirac point), with negligible relaxation during the pulse. In these conditions, the non-equilibrium electronic distribution offers a large phase space for impact ionization \cite{Winzer_2010, Winzer_2012, Plötzing_2014, Gierz_2015_10fs}, which leads to a rapid accumulation of charge carriers at the bottom of the conduction band (Fig. \ref{fig1}a).
  
For longer pump pulses of durations of severals tens of femtoseconds, ultrafast electron-electron scattering events quickly redistribute the photo-excited electrons at E$_{\text{D}}+\hbar\omega_{\text{pump}}/2$ and holes at E$_{\text{D}}-\hbar\omega_{\text{pump}}/2$, allowing for a much larger number of carriers to be excited. Once these carriers reach the Dirac point, they presumably relax by Auger recombination (Fig. \ref{fig1}b) \cite{Winzer_2013, Gierz_2013, Gierz_2015_popinv} and phonon emission \cite{Kampfrath_2005, Yan_2009, Kang_2010, Chatzakis_2011, Johannsen_2013}.

\section{Experimental Methods}

Time- and angle-resolved photoemission experiments have been performed at Artemis at the Central Laser Facility in Harwell, UK. The tr-ARPES setup is based on a 30fs-1kHz Titanium:Sapphire (Ti:Sa) laser system. Extreme ultraviolet (XUV) probe pulses were generated by high order harmonics generation (HHG) in argon. From the resulting broad XUV spectrum $\hbar\omega_{\text{probe}}$=30eV was selected with a time-preserving grating monochromator \cite{Frassetto_2011}. Near-infrared pump pulses at $\hbar\omega_{\text{pump}}$=950meV were generated with an optical parametric amplifier. The cross correlation between the 950meV pump and the 30eV probe pulses was $\sigma$=35fs (FWHM=80fs), determined from the width of the rising edge of the pump probe signal. In order to improve the temporal resolution of the tr-ARPES experiment, $<$10fs infrared pulses were generated by broadening the output spectrum of the Ti:Sa laser system in a neon filled fiber and compressing the pulses using chirped mirrors \cite{Gierz_2015_10fs}. These pulses were then used both as a pump and for HHG resulting in a temporal resolution of $\sigma$=9fs (FWHM=21fs).

Hydrogen-intercalated monolayer graphene samples were grown on 6H-SiC(0001) as described previously \cite{Riedl_2009}. Due to charge transfer from the substrate the graphene layer is hole-doped with the chemical potential $\sim$200meV below the Dirac point.

\section{Results}

\begin{figure*}
	\center
  \includegraphics[width = 1\columnwidth]{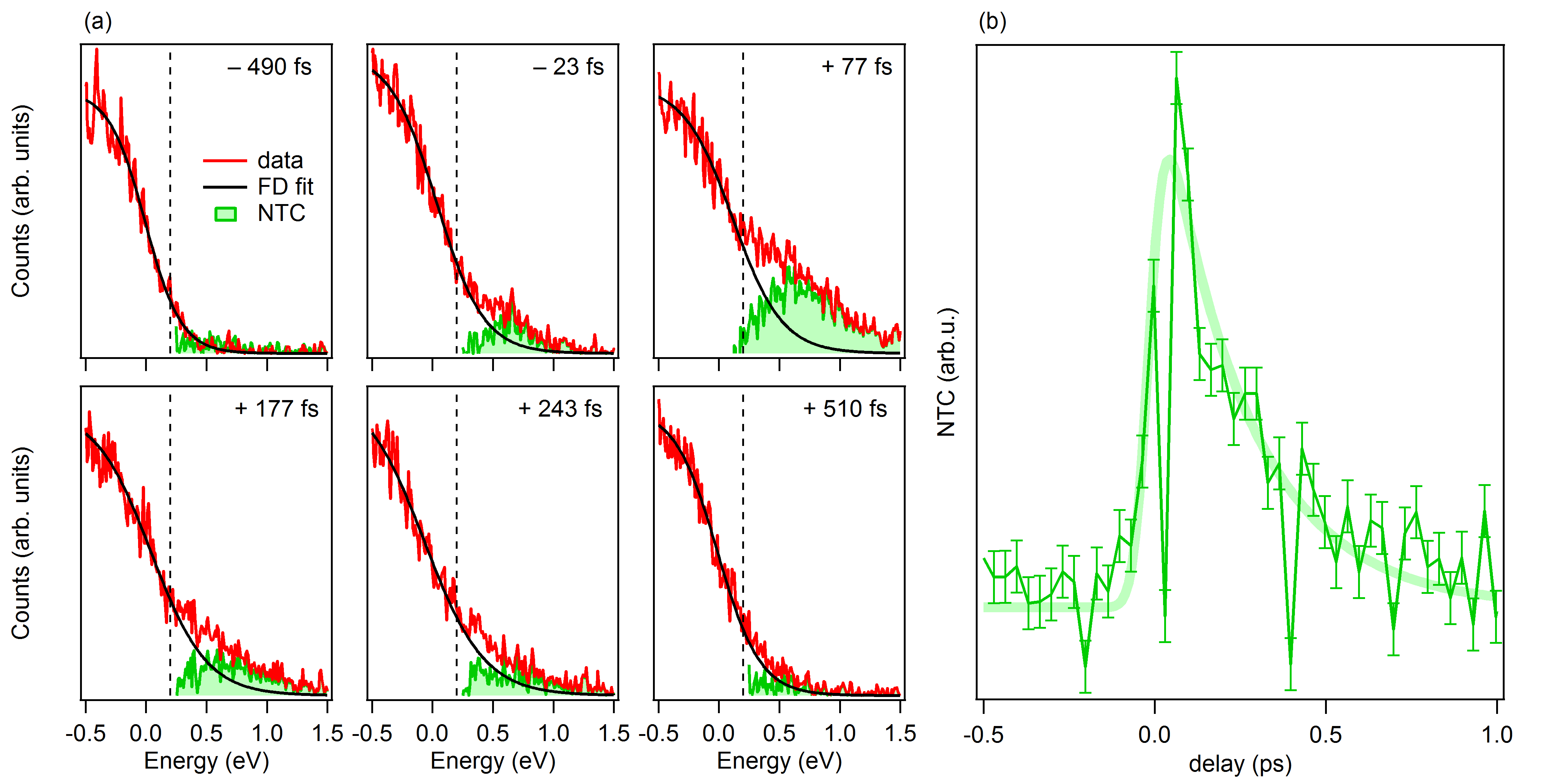}
  \caption{Carrier dynamics for excitation at $\hbar\omega_{\text{pump}}$=950meV probed with a temporal resolution of $\sigma$=35fs (FWHM=80fs). (a) Momentum-integrated electron distribution functions in the vicinity of the K-point (red) together with Fermi-Dirac distribution fits (black). The deviation between fit and data (green) indicates the presence of non-thermal carriers. (b) Number of non-thermal carriers (green area from a) as a function of pump-probe delay. The NTC lifetime as obtained from an exponential fit to the data (dark green line) is 250fs.}
  \label{fig2}
\end{figure*}

Figure \ref{fig2}a shows electronic distribution functions measured with a temporal resolution of $\sigma$=35fs for various pump-probe delays before, during, and after photo-excitation at $\hbar\omega_{\text{pump}}$=950meV. We measured the photocurrent along the $\Gamma$K-direction in the vicinity of the K-point of the hexagonal Brillouin zone. Due to the linear slope of the $\pi$-bands, electronic distribution functions can be simply obtained by integrating the photocurrent over momentum \cite{Gierz_2013, Ulstrup_2014, Gierz_2015_popinv}. At negative delay the data can be nicely fitted with a Fermi-Dirac distribution indicating a thermal carrier distribution. In the presence of the pump pulse a strong bump develops above the Dirac point (dashed line in Fig. \ref{fig2}a) indicating the presence of non-thermal carriers (NTCs). The number of NTCs as a function of pump-probe delay, determined from the area of the residual between Fermi-Dirac fit and data, is shown in Fig. \ref{fig2}b. From an exponential fit to the data we obtain a relaxation time of the NTCs of 250fs.

\begin{figure*}
	\center
  \includegraphics[width = 1\columnwidth]{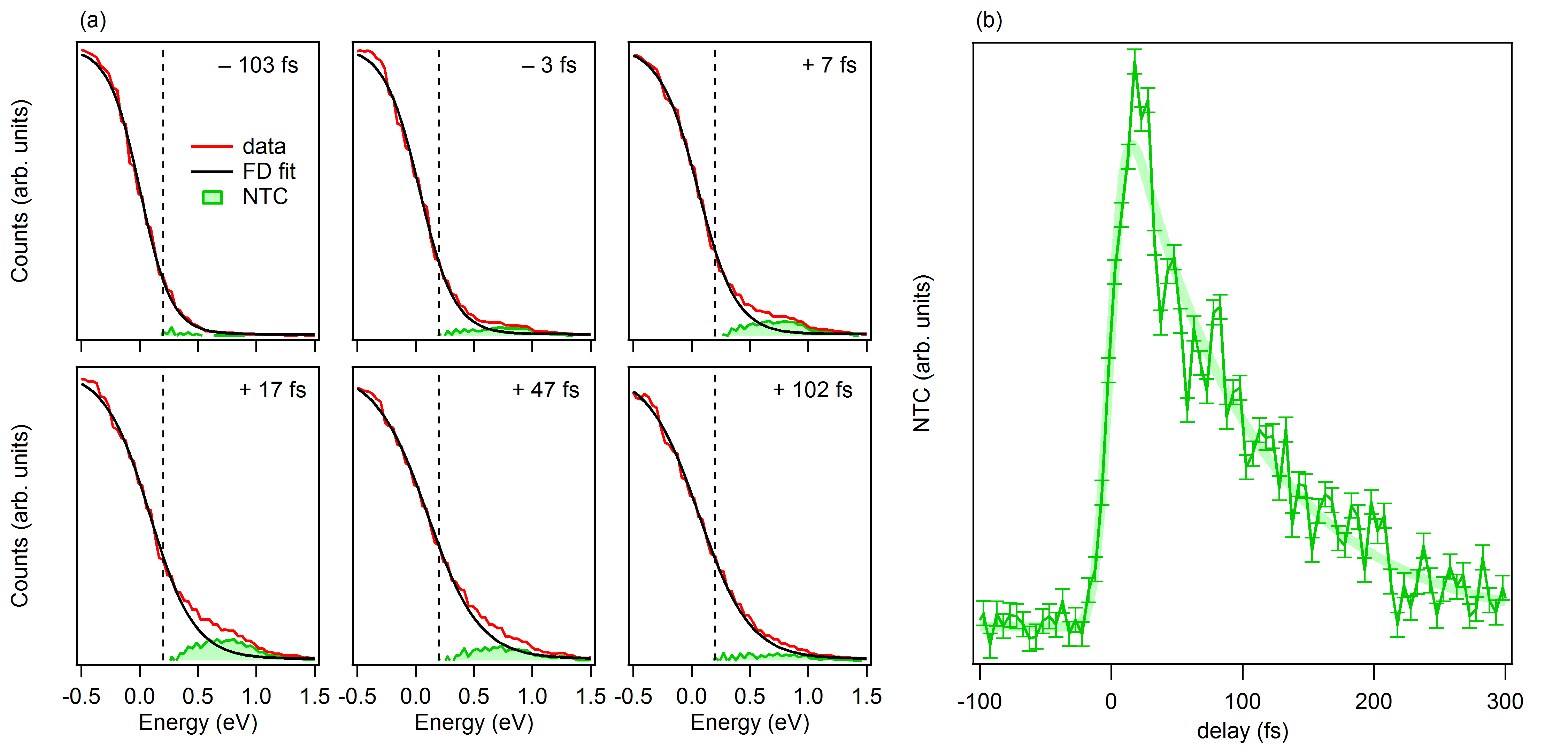}
  \caption{Same as Fig. \ref{fig2} but for $\hbar\omega_{\text{pump}}$=1.6eV and a temporal resolution of $\sigma$=9fs (FWHM=21fs). Compared to Fig. \ref{fig2} the amount of NTCs is smaller and the NTC lifetime is shorter (95fs).}
  \label{fig3}
\end{figure*}

When using $<$10fs pump pulses at $\hbar\omega_{\text{pump}}$=1.6eV instead (see Fig. \ref{fig3}), the maximum number of NTCs is smaller and their lifetime shorter (95fs) compared to excitation with long pulses.

\begin{figure*}
	\center
  \includegraphics[width = 1\columnwidth]{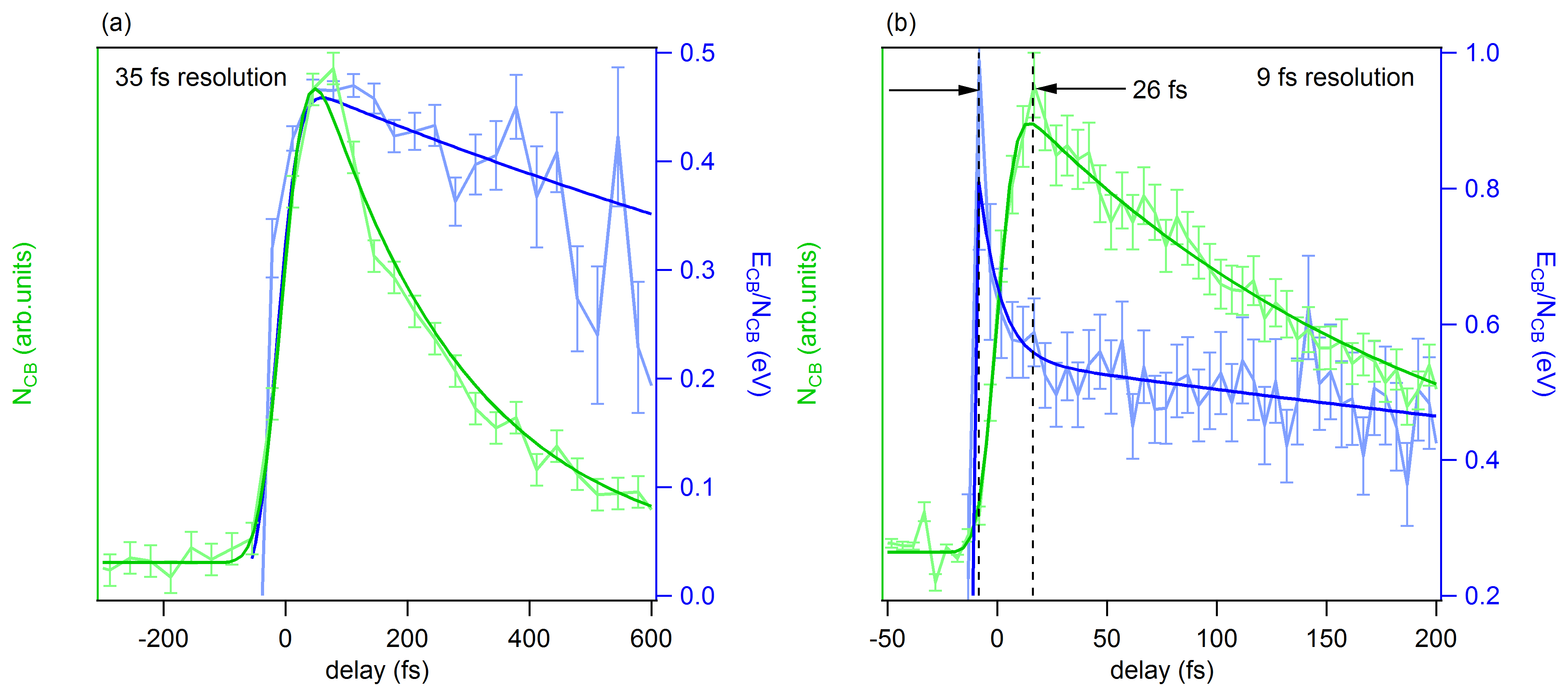}
  \caption{Total number of carriers inside the conduction band, N$_{\text{CB}}$, and their average kinetic energy, E$_{\text{CB}}$/N$_{\text{CB}}$, as a function of pump probe delay for a temporal resolution of $\sigma$=35fs (a) and 9fs (b). The better temporal resolution in (b) allows for the identification of impact ionization as the primary thermalization event on femtosecond time scales.}
  \label{fig4}
\end{figure*}

In Fig. \ref{fig4} we take a closer look at the primary scattering events that contribute to the buildup of the NTC distribution at the bottom of the conduction band. The ultrafast few-femtosecond dynamics in graphene are expected to be dominated by different Auger processes, in particular by Auger recombination and impact ionization \cite{Winzer_2010, Winzer_2012}. These processes can be easiliy distinguished by measuring the number of carriers inside the conduction band, N$_{\text{CB}}$, as well as their average kinetic energy, E$_{\text{CB}}$/N$_{\text{CB}}$, in a tr-ARPES experiment \cite{Gierz_2015_10fs}. Auger recombination (Fig. \ref{fig1}b) decreases N$_{\text{CB}}$ and increases E$_{\text{CB}}$/N$_{\text{CB}}$. The inverse process, impact ionization (Fig. \ref{fig1}a), increases N$_{\text{CB}}$ and reduces E$_{\text{CB}}$/N$_{\text{CB}}$. The available phase space dictates which one of these two processes dominates the carrier relaxation \cite{Winzer_2010, Winzer_2012}. With a temporal resolution of $\sigma$=35fs (Fig. \ref{fig4}a), N$_{\text{CB}}$ and E$_{\text{CB}}$/N$_{\text{CB}}$ seem to increase with the same time constant making a distinction between Auger recombination and impact ionization impossible. Thus, we repeated the measurement with a better temporal resolution of $\sigma$=9fs (Fig. \ref{fig4}b) \cite{Gierz_2015_10fs}. For the first 25fs after photo-excitation, we observe an increase of N$_{\text{CB}}$ while E$_{\text{CB}}$/N$_{\text{CB}}$ is already decreasing, clearly indicating the dominance of impact ionization over Auger recombination.

\section{Discussion}
 
From Fig. \ref{fig4} it is obvious that Auger scattering in photo-excited graphene during the early stages of carrier relaxation is extremely fast, $\leq$25fs. Impact ionization rapidly accumulates carriers at the bottom of the conduction band, assisting the buildup of the observed NTC distribution as illustrated in Fig. \ref{fig1}a. The total number of NTCs in the present experiment is determined by the pump pulse duration. The employed pump fluences (20mJ/cm$^2$ for the short pulses, 4.6mJ/cm$^2$ for the long pulses) are in the regime where the absorption saturates \cite{Winzer_2012_II, Winnerl_2013}. In this case, we expect a higher number of excited carriers for long pump pulses, because the states that are depleted by the pump pulse are constantly refilled by electron-electron scattering while the pump pulse is still on.

The decay mechansim for the NTC distribution is sketched in Fig. \ref{fig1}b. Once the carriers have accumulated at the bottom of the conduction band, their relaxation is dominated by Auger recombination with a minor contribution of optical phonon emission \cite{Winzer_2013}. Compared to the ultrafast buildup of the NTC distribution via impact ionization on timescales shorter than 25fs, its lifetime of 250fs and 95fs for long and short pump pulses, respectively, seems long. This can be understood as follows. Once the carriers have accumulated close to the Dirac point where the density of states and thus the available scattering phase space is small, Auger scattering is slowed down resulting in the observed timescale on the order of 100fs. Further, the lifetime of the NTC distribution is found to depend on the number of NTCs that have accumulated at the Dirac point. This indicates that the carriers pass through the Dirac cone at a constant rate.

\section{Conclusion}

In summary, we have tracked the buildup and decay of non-thermal carriers after optical interband excitation in monolayer graphene with tr-ARPES. Using $<$10fs pulses we have identified impact ionization as the primary scattering channel during the first 25fs after photo-excitation \cite{Gierz_2015_10fs}. Once the carriers have accumulated at the bottom of the conduction band they relax via Auger recombination \cite{Winzer_2013, Gierz_2013, Gierz_2015_popinv}. The lifetime of the NTC distribution was found to be on the order of 100fs, with longer lifetimes for a higher number of NTCs, indicating a constant relaxation rate through the Dirac point. The number of NTCs in the present high fluence regime is determined by the pump pulse duration. For long pump pulses (several tens of femtoseconds) the excited electrons are redistributed by ultrafast electron-electron scattering wihtin the pulse duration, reducing absorption bleaching and allowing for a much bigger number of photo-excited carriers compared to short pulses ($<$10fs).
 
High fluence excitation with long pump pulses, the same as the one in Fig. \ref{fig2},  has been shown to result in a population-inverted state with potential applications in TeraHertz lasing \cite{Gierz_2013, Gierz_2015_popinv}. Further, impact ionization in principle allows for efficient light harvesting, as the absorption of a single photon may generate multiple electron-hole pairs \cite{Kolodinski_1993, Winzer_2010, Winzer_2012, Plötzing_2014, Gierz_2015_10fs}. However, the absence of a band gap in graphene makes the separation of electrons and holes difficult and solar cell applications unlikely. 

\section{Acknowledgments} 

The research leading to these results has received funding from LASERLAB-EUROPE (grant agreement no. 284464, EC's Seventh Framework Programme), the Science and Technology Facilities Council (STFC), and the Priority Program 1459 ``Graphene'' of the German Science Foundation. We thank Francesca Calegari, Cephise Cacho, Sven Aeschlimann, and Matteo Mitrano who made the experiments possible. Further, we thank Andrea Cavalleri for many lively discussions and for his input on the manuscript. We thank Ermin Mali{\'c} for many helpful discussions.

\clearpage

\begin{thebibliography}{12}

\bibitem{book} J. T. Verdeyen, Laser Electronics, third edition (1995)
\bibitem{Mahan_1979} J. E. Mahan, T. W. Ekstedt, R. I. Frank, and R. Kaplow, IEEE Transactions on Electron Devices 26, 733 (1979)
\bibitem{Palais_2000} O. Palais, J. Gervais, L. Clerc, S. Martinuzzi, Materials Science and Engineering B71, 47 (2000)
\bibitem{Cuevas_2003} A. Cuevas, and D. Macdonald, Solar Energy 76, 255 (2004)
\bibitem{Fuyuki_2005} T. Fuyuki, H. Kondo, T. Yamazaki, Y. Takahashi, and Y. Uraoka, Appl. Phys. Lett. 86, 262108 (2005)
\bibitem{Schmitt_2008} F. Schmitt, P. S. Kirchmann, U. Bovensiepen, R. G. Moore, L. Rettig, M. Krenz,
J.-H. Chu, N. Ru, L. Perfetti, D. H. Lu, M. Wolf, I. R. Fisher, and Z.-X. Shen, Science 321, 1649  (2008)
\bibitem{Rohwer_2011} T. Rohwer, S. Hellmann, M. Wiesenmayer, C. Sohrt, A. Stange, B. Slomski, A. Carr, Y. Liu, L. M. Avila, M. Kalläne, S. Mathias, L. Kipp, K. Rossnagel, and M. Bauer, Nature 471, 490 (2011) 
\bibitem{Petersen_2011} J. C. Petersen, S. Kaiser, N. Dean, A. Simoncig, H.Y. Liu, A. L. Cavalieri, C. Cacho, I. C. E. Turcu, E. Springate, F. Frassetto, L. Poletto, S. S. Dhesi, H. Berger, and A. Cavalleri, Phys. Rev. Lett. 107, 177402 (2011) 
\bibitem{Smallwood_2012} C. L. Smallwood, J. P. Hinton, C. Jozwiak, W. Zhang, J. D. Koralek, H. Eisaki, D.-H. Lee, J. Orenstein, and A. Lanzara, Science 336, 1137 (2012) 
\bibitem{Bovensiepen_2012} U. Bovensiepen, and P. S. Kirchmann, Laser Photonics Rev. 6, 589 (2012) 
\bibitem{Hellmann_2012} S. Hellmann, T. Rohwer, M. Kalläne, K. Hanff, C. Sohrt, A. Stange, A. Carr, M. M. Murnane, H. C. Kapteyn, L. Kipp, M. Bauer, and K. Rossnagel, Nat. Commun. 3, 1069 (2012) 
\bibitem{Sentef_2013} M. Sentef, A. F. Kemper, B. Moritz, J. K. Freericks, Z.-X. Shen, and T. P. Devereaux, Phys. Rev. X 3, 041033 (2013) 
\bibitem{Novoselov_2004} K. S. Novoselov, A. K. Geim, S. V. Morozov, D. Jiang, Y. Zhang, S. V. Dubonos, I. V. Grigorieva, and A. A. Firsov, Science 306, 666 (2004) 
\bibitem{Bostwick_2007} A. Bostwick, T. Ohta, T. Seyller, K. Horn and E. Rotenberg, Nat. Phys. 3, 36 (2007) 
\bibitem{Winzer_2013} T. Winzer, E. Mali{\'c}, and A. Knorr, Phys. Rev. B 87, 165413 (2013) 
\bibitem{Ryzhii_2007} V. Ryzhii, M. Ryzhii, and T. Otsuji, Appl. Phys. Lett. 101, 083114 (2007) 
\bibitem{Li_2012} T. Li, L. Luo, M. Hupalo, J. Zhang, M. C. Tringides, J. Schmalian, and J. Wang, Phys. Rev. Lett. 108, 167401 (2012) 
\bibitem{Gierz_2013} I. Gierz, J. C. Petersen, M. Mitrano, C. Cacho, E. Turcu, E. Springate, A. St{\"o}hr, A. K{\"o}hler, U. Starke, and A. Cavalleri, Nat. Mater. 12, 1119 (2013)
\bibitem{Gierz_2015_popinv} I. Gierz, M. Mitrano, J. C. Petersen, C. Cacho, I. C. E. Turcu, E. Springate, A. St{\"o}hr, A. K{\"o}hler, U. Starke, and A. Cavalleri. J. Phys.: Condens. Matter 27, 164204 (2015)
\bibitem{Jago_2015} R. Jago, T. Winzer, A. Knorr, and E. Mali{\'c}, Phys. Rev. B 92, 085407 (2015) 
\bibitem{Beattie_1958} A. R. Beattie, and P. T. Landsberg, Proc. R. Soc. A 249, 16 (1958) 
\bibitem{Svantesson_1971} K. G. Svantesson, N. G. Nilsson, and L. Huldt, Solid State Commun. 9, 213 (1971) 
\bibitem{Auston_1975} D. H. Auston, C. V. Shank, and P. LeFur, Phys. Rev. Lett. 35, 1022 (1975) 
\bibitem{Benz_1976} G. Benz, and R. Conradt, Phys. Rev. B 16, 843 (1976)
\bibitem{Winzer_2010} T. Winzer, A. Knorr, and E. Mali{\'c}, Nano Lett. 10, 4839 (2010)
\bibitem{Winzer_2012} T. Winzer, and E. Mali{\'c}, Phys. Rev.  B 85, 241404(R) (2012) 
\bibitem{Plötzing_2014} T. Pl{\"o}tzing, T. Winzer, E. Mali{\'c}, D. Neumaier, A. Knorr, and H. Kurz, Nano Lett. 14, 5371 (2014) 
\bibitem{Gierz_2015_10fs} I. Gierz, F. Calegari, S. Aeschlimann, M. Chavez Cervantes, C. Cacho, R. T. Chapman, E. Springate, S. Link, U. Starke, C. R. Ast, and A. Cavalleri, Phys. Rev. Lett. 115, 086803 (2015) 
\bibitem{Kampfrath_2005} T. Kampfrath, L. Perfetti, F. Schapper, C. Frischkorn, and M. Wolf, Phys. Rev. Lett. 95, 187403 (2005) 
\bibitem{Yan_2009} H. Yan, D. Song, K. F. Mak, I. Chatzakis, J. Maultzsch, and T. F. Heinz, Phys. Rev. B 80, 121403(R) (2009) 
\bibitem{Kang_2010} K. Kang, D. Abdula, D. G. Cahill, and M. Shim, Phys. Rev. B 81, 165405 (2010) 
\bibitem{Chatzakis_2011} I. Chatzakis, H. Yan, D. Song, S. Berciaud, and T. F. Heinz, Phys. Rev. B 83, 205411 (2011)
\bibitem{Johannsen_2013} J. C. Johannsen, S. Ulstrup, F. Cilento, A. Crepaldi, M. Zacchigna, C. Cacho, I. C. E. Turcu, E. Springate, F. Fromm, C. Raidel, T. Seyller, F. Parmigiani, M. Grioni, and P. Hofmann, Phys. Rev. Lett. 111, 027403 (2013)
\bibitem{Frassetto_2011} F. Frassetto, C. Cacho, C. A. Froud, I. C. E. Turcu, P. Villoresi, W. A. Bryan, E. Springate, and L. Poletto, Opt. Express 19, 19169 (2011)
\bibitem{Riedl_2009} C. Riedl, C. Coletti, T. Iwasaki, A. A. Zakharov, and U. Starke, Phys. Rev. Lett. 103, 246804 (2009)
\bibitem{Ulstrup_2014} S. Ulstrup, J. C. Johannsen, M. Grioni, and P. Hofmann, Rev. Sci. Instr. 85, 013907 (2014)
\bibitem{Winzer_2012_II} T. Winzer, A. Knorr, M. Mittendorff, S. Winnerl, M.-B. Lien, D. Sun, T. B. Norris, M. Helm, and E. Mali{\'c}, Appl. Phys. Lett. 101, 221115 (2012)
\bibitem{Winnerl_2013} S. Winnerl, F. G{\"o}ttfert, M. Mittendorff, H. Schneider, M. Helm, T. Winzer, E. Mali{\'c}, A. Knorr, M. Orlita, M. Potemski, M. Sprinkle, C. Berger, and W. A. de Heer, J. Phys.: Condens. Matter 25, 054202 (2013)
\bibitem{Kolodinski_1993} S. Kolodinski, J. H. Werner, T. Wittchen, and H. J. Queisser, Appl. Phys. Lett. 63, 2405 (1993)

\end{thebibliography}
\end{document}